%% file: HasanOgul_dpf2015.tex
\documentclass[12pt]{article}
\usepackage{graphicx}

\newcommand\pubnumber{DPF2015-31}
\newcommand\pubdate{\today}

\def\iowa{The University of Iowa\\
Iowa City, Iowa, USA}
\def\support{\footnote{on behalf of the CMS Collaboration.}}

\def\Title#1{\begin{center} {\Large #1 } \end{center}}
\def\Author#1{\begin{center}{ \sc #1} \end{center}}
\def\Address#1{\begin{center}{ \it #1} \end{center}}

\newcommand\pubblock{\rightline{\begin{tabular}{l} \pubnumber\\
         \pubdate  \end{tabular}}}
\newenvironment{Abstract}{\begin{quotation}  }{\end{quotation}}
\newenvironment{Presented}{\begin{quotation} \begin{center} 
             PRESENTED AT\end{center}\bigskip 
      \begin{center}\begin{large}}{\end{large}\end{center} \end{quotation}}
\def\Acknowledgments{\bigskip  \bigskip \begin{center} \begin{large}
             \bf ACKNOWLEDGMENTS \end{large}\end{center}}
\input econfmacros.tex
\begin{document}
\begin{titlepage}
\pubblock

\vfill
\Title{Measurement of the Muon Charge Asymmetry for W Bosons Produced in Inclusive $pp\rightarrow W(\mu\nu) + X$ at $\sqrt{s}$ = 8 TeV}
\vfill
\Author{Hasan Ogul\support}
%\Author{Hasan Ogul on behalf of the CMS Collaboration}
\Address{\iowa}
\vfill
\begin{Abstract}
Measurement of the muon charge asymmetry in inclusive $pp$$\rightarrow$$W(\mu\nu)$ + X at $\sqrt{s}$ = 8 TeV is presented. The data sample corresponds to an integrated luminosity of 18.8 $fb^{-1}$ recorded with the CMS detector at the LHC. With a sample of more than a hundred million $W$$\rightarrow$$\mu$$\nu$ events, the statistical precision is greatly improved in comparison to previous measurements. This new result can provide additional constraints on the parton distribution functions of the proton. This measurement is used together with the cross sections for inclusive deep inelastic ep scattering at HERA in a next-to-leading-order QCD analysis. The impact to the valence quark distributions is demonstrated.
\end{Abstract}
\vfill
\begin{Presented}
DPF 2015\\
The Meeting of the American Physical Society\\
Division of Particles and Fields\\
Ann Arbor, Michigan, August 4--8, 2015\\
\end{Presented}
\vfill
\end{titlepage}
\def\thefootnote{\fnsymbol{footnote}}
\setcounter{footnote}{0}

\section{Introduction}

In $pp$ collisions, the primary mechanism of the W production is mainly from $u\bar{d}$$\rightarrow$$W^{+}$ and $d$$\bar{u}$$\rightarrow$$W^{-}$. Since the number of $u$ valance quarks are greater than the number of $d$ valance quarks in the proton, $W^{+}$ bosons occur more frequently than $W^{-}$ in $pp$ collisions. This difference between $W^{+}$ and $W^{-}$ bosons motivates a study called W boson charge asymmetry. The measurement of the asymmetry in the production of $W^{+}$ and $W^{-}$ bosons as a function of boson rapidity is sensitive to $d/u$ ratio and the sea quark densities in the proton. However, it is difficult to measure the W boson production asymmetry as a function of the boson rapidity because of the presence of neutrinos in the leptonic W decay. Therefore, the muon charge asymmetry in terms of muon pseudorapidity ($\eta$) is measured, which is defined by Eq.~(\ref{eq:asym}). 

\begin{equation}
\label{eq:asym}
A(\eta) = \frac{N^{W^{+}(\mu\nu)}-N^{W^{-}(\mu\nu)}}{N^{W^{+}(\mu\nu)}+N^{W^{-}(\mu\nu)}}
\end{equation}

The measurement of this asymmetry in the colliding protons can allow us to test higher order calculations and provide new insights into proton structure~\cite{PDF}. Therefore, any precise measurement of the charge asymmetry can contribute to improvements of theoretical predictions that rely on precise knowledge of the PDFs. 

\section{Event Selection and Background Discussion}
The Compact Muon Solenoid (CMS) experiment is one of the general-purpose experiments at the Large Hadron Collider (LHC)~\cite{cmsTDR}. Recently, CMS has performed a measurement of the muon charge asymmetry as a function of muon pseudorapidity in inclusive $W$$\rightarrow$$\mu\nu$ production at $\sqrt{s}$$=$8 TeV~\cite{Cam8TeV}. The $W$$\rightarrow$$\mu\nu$ events were collected by a single isolated muon trigger which requires muon $p_{T}$$>$24 GeV and $|\eta|$$<$2.4. In order to suppress background contamination, identification and isolation criteria are applied on the reconstructed muons. Then, the muons in each event are sorted by $p_{T}$ and the leading one is taken as $W(\mu\nu)$ candidate. The leading muon is further required to have $p_{T}$$>$25 GeV. In order to reduce the Drell-Yan background, events with an additional muon with $p_{T}$$>$15 GeV are rejected. The detailed information can be found in reference~\cite{Cam8TeV}. \par
A total of 61 million $W^{+}$$\rightarrow$$\mu^{+}\nu$ and 45 million $W^{-}$$\rightarrow$$\mu^{-}\nu$ candidate events are selected. After the selection of events, the asymmetry is measured in 11 bins of absolute muon pseudorapidity. The bins are:
\begin{center}
[0, 0.2, 0.4, 0.6, 0.8, 1.0, 1.2, 1.4, 1.6, 1.85, 2.1, 2.4]
\end{center}
There are some dominant processes contributing to the total backgrounds:
\begin{itemize}
\item Multijet (QCD) events with high $p_{T}$ muons
\item Drell-Yan events decaying into $\mu^{+}$ and $\mu^{-}$
\item Electro-­weak, including Drell-Yan decaying into $\tau^{+}\tau^{-}$ and $W(\tau\nu)$, and $t\bar{t}$ events
\end{itemize}

\section{Results}
\begin{figure}[htb]
\centering
\includegraphics[height=2.5in]{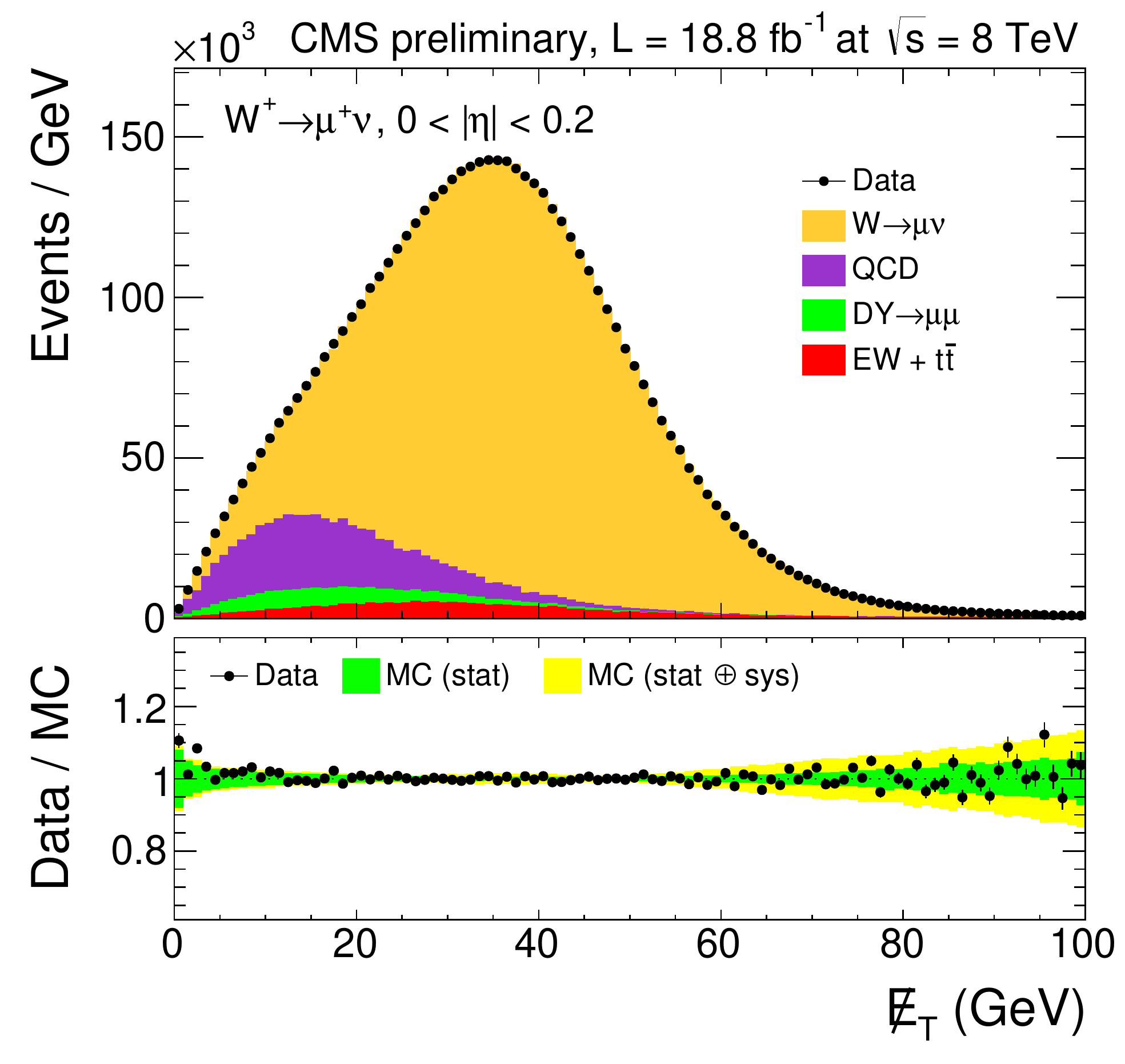}
\includegraphics[height=2.5in]{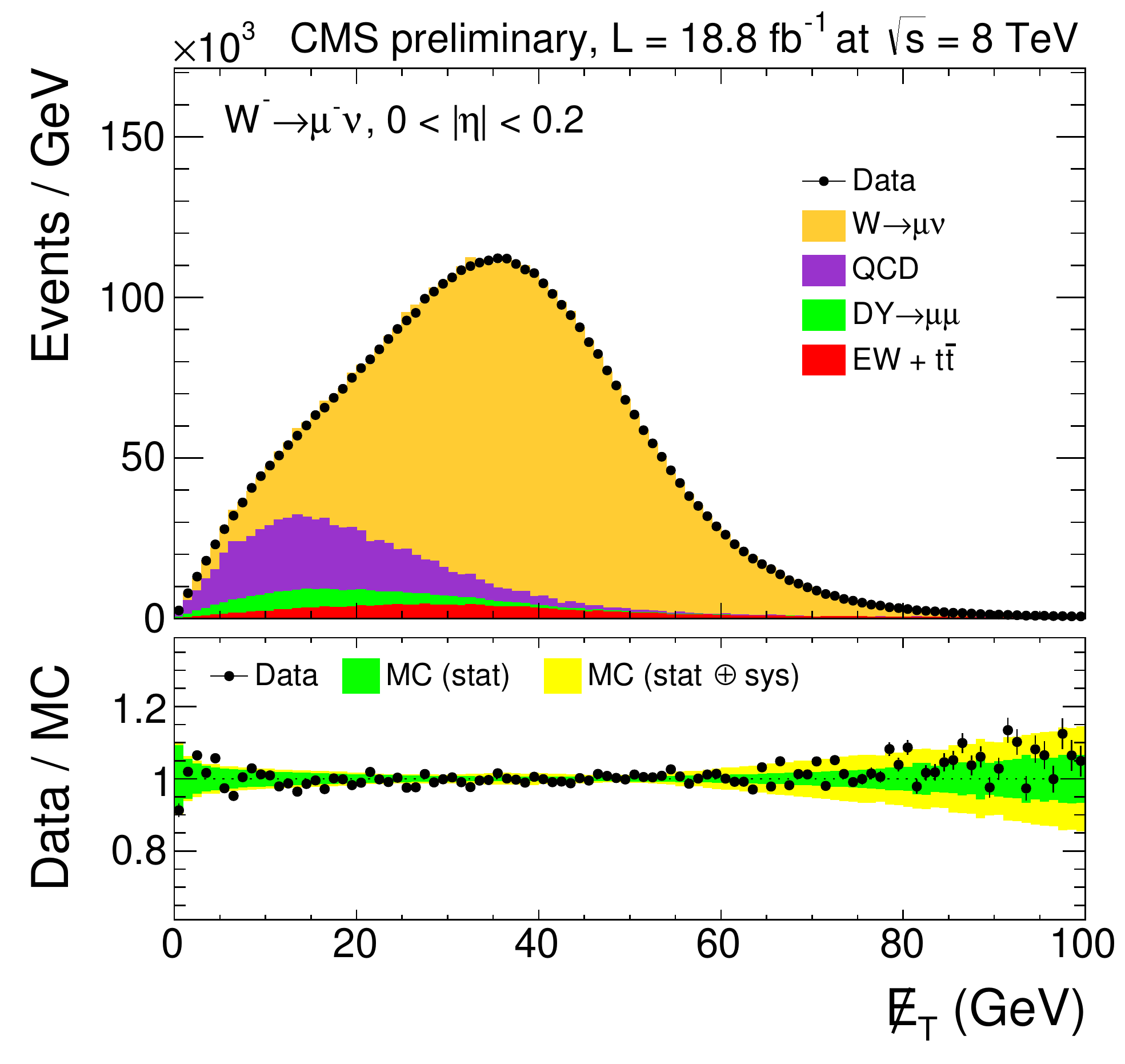}
\caption{The fit plots of $W^{+}$$(\mu\nu)$ and $W^{-}$$(\mu\nu)$ are shown on left and right, respectively.}
\label{fig:fits}
\end{figure}

Numbers of $W^{+}$$\rightarrow$$\mu^{+}$$\nu$ and $W^{-}$$\rightarrow$$\mu^{-}$$\nu$ are extracted by simultaneous fits of missing transverse energy (MET) distributions. In each absolute pseudorapidity bin, MET templates derived from Monte Carlo (MC) simulation samples are fitted to the MET distribution in data for signal and background extraction. The MET templates are further calibrated using $Z(\mu^{+}\mu^{-})$ events~\cite{recoil}.  Figure~\ref{fig:fits} shows the examples of MET fits for the first absolute eta bin. The ratios between MC and data histograms are shown on the lower panels, and it is seen that there is a good agreement of fits with data. \par

\begin{figure}[htb]
\centering
\includegraphics[height=3.5in]{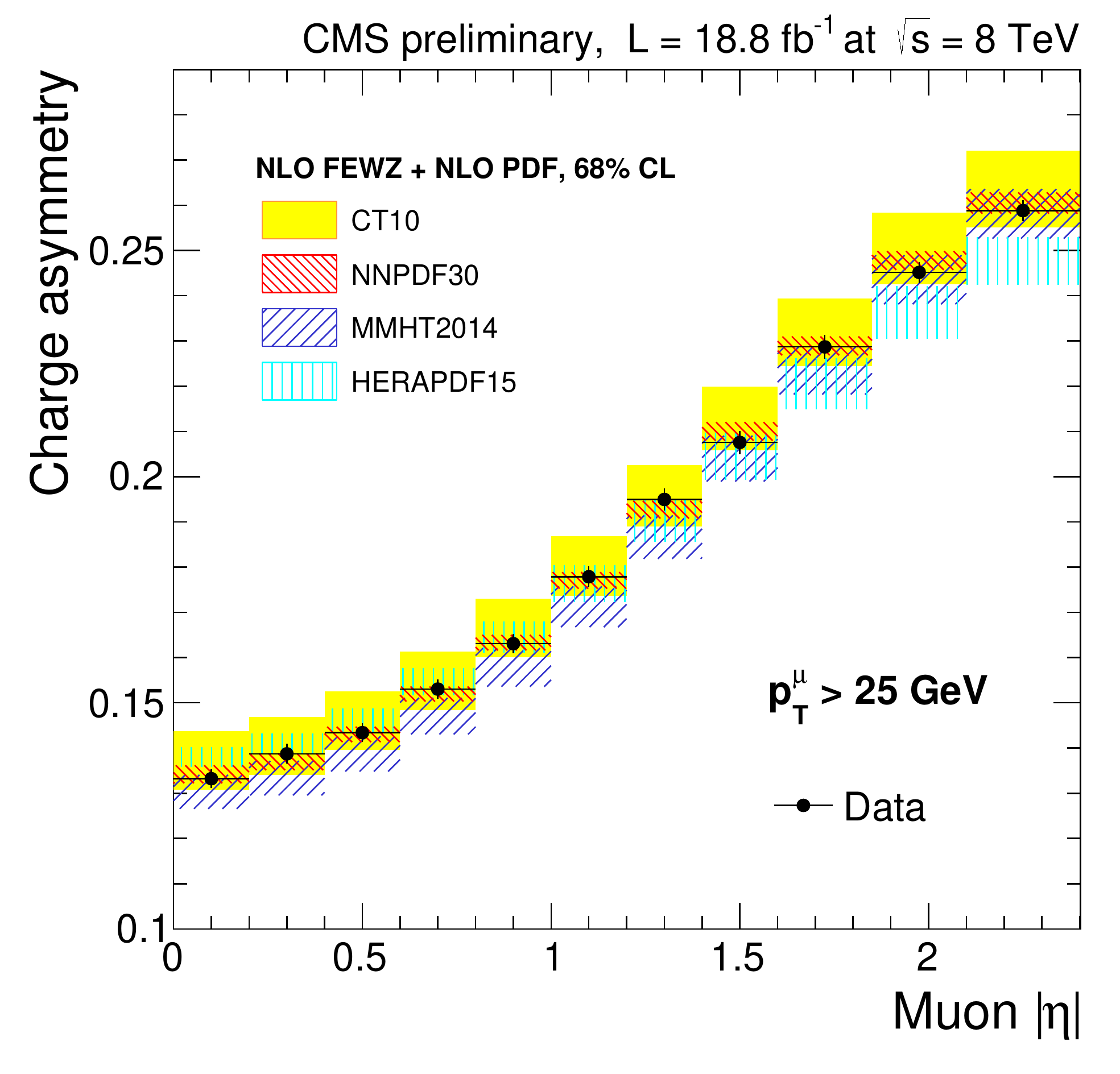}
\caption{Comparison of the measured charge asymmetries with NLO predictions calculated using the FEWZ 3.1.}
\label{fig:Asym}
\end{figure}

The extracted number of $W(\mu\nu)$ signal events in each $|\eta|$ bin are used to calculate the muon charge asymmetry, as defined by Eq.~(\ref{eq:asym}). In addition, the bias to the asymmetry due to difference between $\mu^{+}$ and $\mu^{-}$ selection efficiencies is corrected for, using efficiency values determined in Drell-Yan events~\cite{muonperf}. Figure~\ref{fig:Asym} presents the comparison of the measured asymmetry with theoretical NLO QCD calculations based on 4 different PDF models. The theoretical predictions are obtained using the FEWZ 3.1~\cite{FEWZ} interfaced with CT10~\cite{PDF}, NNPDF3.0~\cite{nnpdf}, HERAPDF1.5~\cite{hera} and MMHT2014~\cite{mmht} PDF sets. The colored bands represent the PDF uncertainties. The data is shown in black and the error bars on the experimental results include systematic and statistical uncertainties. The measurements are compatible with theoretical predictions within their uncertainties. The experimental uncertainties are significantly smaller than PDF uncertainties.

\section{QCD Analysis}
\begin{figure}[htb]
\centering
\includegraphics[height=2.5in]{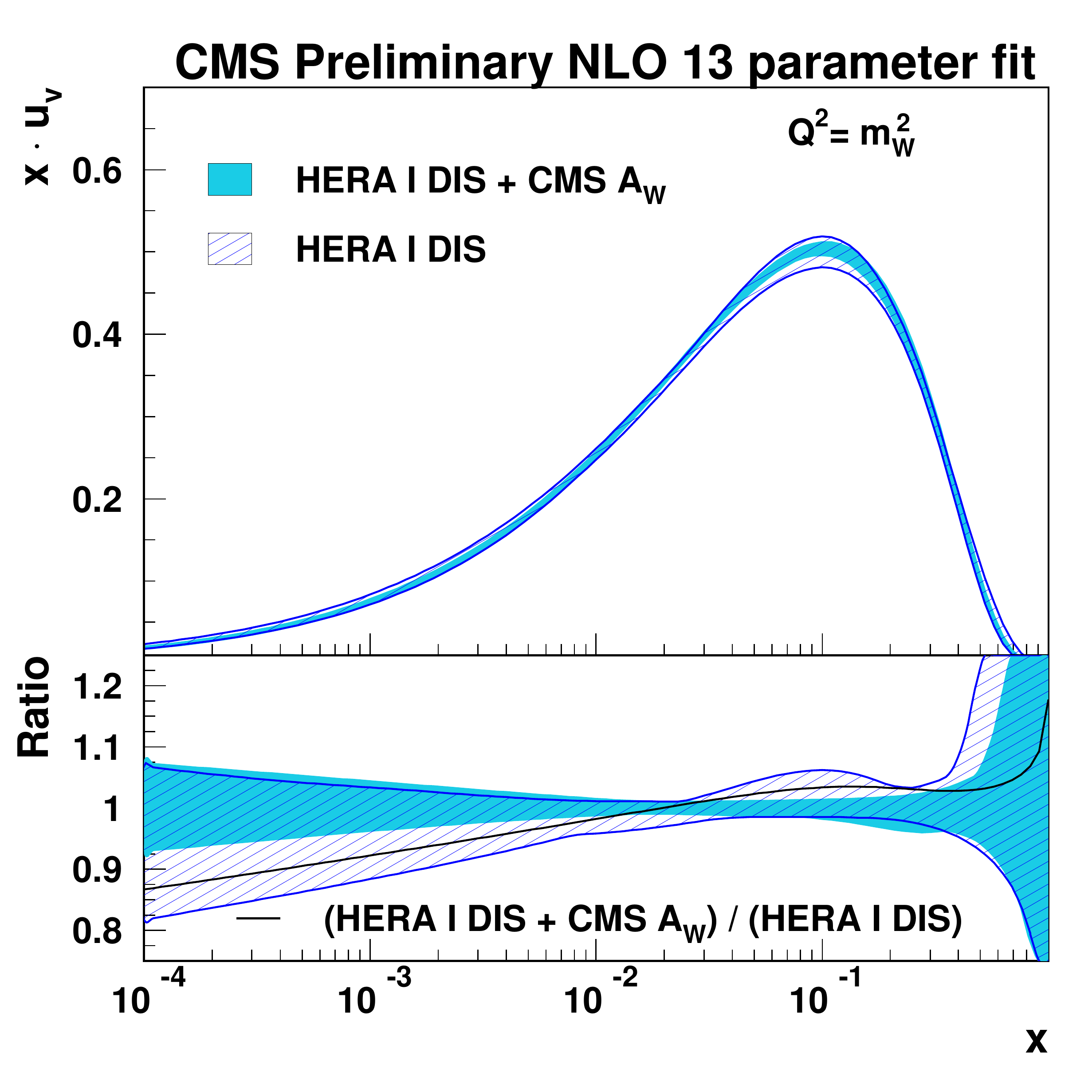}
\includegraphics[height=2.5in]{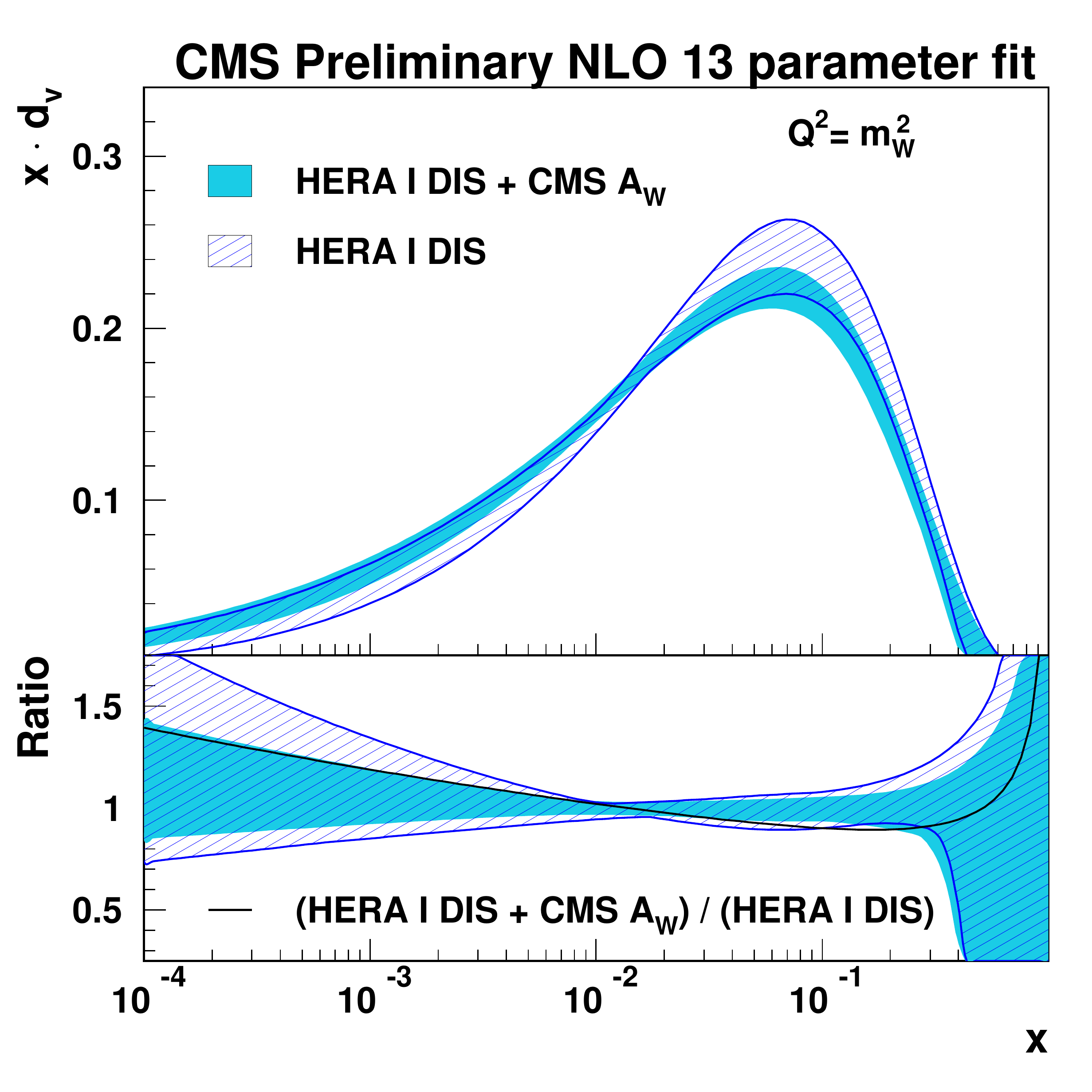}
\caption{ Distributions of u valence (left) and d valence (right) quarks as functions of x at the scale $Q^{2}$=$m^{2}_{W}$.}
\label{fig:QCD}
\end{figure}

To understand the impact of the charge asymmetry measurement on PDFs, the charge asymmetry results are used in a QCD analysis at NLO together with the combined HERA data~\cite{QCD}. The details of the QCD analysis can be found in reference~\cite{Cam8TeV}. Figure~\ref{fig:QCD} shows the impact of the charge asymmetry on $u$ and $d$ valance quark PDF shapes. The results of the fit to the HERA data and muon asymmetry measurements (light shaded band),  and to HERA only (dark hatched band) are compared. It is clearly seen that the charge asymmetry measurement reduces the uncertainties on $u$ and $d$ valance quark PDF shapes.
 
\Acknowledgments
I am grateful to work with Ping Tan, Jiyeon Han, Jane Nachtman, Aleko Khukhunaishvili, Arie Bodek, Saranya Ghosh and Katerina Lipka on this crucial analysis.

\end{document}

%% file: econfmacros.tex
\def\beq{\begin{equation}}
\def\eeq#1{\label{#1}\end{equation}}
\def\eeqn{\end{equation}}

\def\beqa{\begin{eqnarray}}
\def\eeqa#1{\label{#1}\end{eqnarray}}
\def\eeqan{\end{eqnarray}}

\let\bar=\overbar

\def\Dslash{\not{\hbox{\kern-4pt $D$}}}
\def\dslash{\not{\hbox{\kern-2pt $\del$}}}

\def\msb{{\bar{\ssstyle M \kern -1pt S}}}